\documentclass[a4paper,11pt]{article}
\usepackage{pos}
\usepackage{amsmath}
\usepackage{subcaption}
\graphicspath{{figures/}}

\title{Thermal photon production in massless boost-invariant gases:
hydrodynamization without thermalization}

\ShortTitle{Thermal photon production in massless boost-invariant gases}

\author*[a,b]{Caio V. P. de Brito}
\author[c]{Gabriel S. Denicol}

\affiliation[a]{Department of Physics, University of Jyväskylä, P.O.B. 35, FI-40014 Jyväskylä, Finland}
\affiliation[b]{Helsinki Institute of Physics, P.O.B. 64, FI-00014 Helsinki, Finland}
\affiliation[c]{Instituto de F\'{\i}sica, Universidade Federal Fluminense, Av.~Gal.~Milton Tavares de Souza, \\ 
S/N, 24210-346, Gragoatá, Niter\'{o}i, Rio de Janeiro, Brazil}

\emailAdd{caio.vp.debrito@jyu.fi}
\emailAdd{gsdenicol@id.uff.br}

\abstract{We study hydrodynamization and thermalization in ultrarelativistic heavy-ion collisions using relativistic kinetic theory. Modelling the quark-gluon plasma as a classical gas of massless particles undergoing a boost-invariant longitudinal expansion, we solve the Boltzmann equation in the relaxation-time approximation via a semi-analytical method based on the method of moments. We show that the energy-momentum tensor is well described by relativistic fluid dynamics even when the system is far from local equilibrium. The single-particle distribution function, on the other hand, is not well approximated by a hydrodynamic truncation, suggesting that the onset of hydrodynamic behaviour does not require thermalization. We then quantify how the departure from equilibrium affects the emission rate of thermal photons in the forward scattering approximation, and show that non-equilibrium corrections beyond the hydrodynamic limit can have a significant impact on this observable.}

\FullConference{
  ICHEP 2026 -- 43rd International Conference on High Energy Physics\\
  Natal, RN, Brazil \\
  July 27 -- August 2, 2026
}

\begin{document}
\maketitle

\section{Introduction}
\label{sec:intro}

Ultrarelativistic heavy-ion collisions provide experimental access to strongly interacting matter under extreme conditions. After two heavy nuclei collide, the system evolves into the quark-gluon plasma (QGP), a state of matter in which quarks and gluons are deconfined. Remarkably, the evolution of this system is well described by relativistic viscous fluid dynamics~\cite{Heinz:2013th}, even when the system is expected to be far from equilibrium. This raises a fundamental question: to what extent does hydrodynamic applicability require proximity to local thermodynamic equilibrium?

The process by which the evolution of the fireball admits a fluid-dynamical description is called \textit{hydrodynamization}. Thermalization, in contrast, refers to the approach to local thermodynamic equilibrium -- which, in the context of kinetic theory, means that the single-particle distribution function $f_{\mathbf{k}}$ approaches a local equilibrium distribution. Understanding whether hydrodynamization implies thermalization is an open problem in the field. In this contribution, we address this question using relativistic kinetic theory. We model the QGP as a classical gas of massless particles undergoing a boost-invariant longitudinal expansion, assuming Bjorken flow \cite{Bjorken:1982qr}, and solve the Boltzmann equation exactly within the relaxation-time approximation (RTA) \cite{Anderson:1974nyl}. By comparing the exact solution to the fluid-dynamical limit, we show that the energy-momentum tensor $T^{\mu\nu}$ is well described by a hydrodynamic truncation whereas $f_{\mathbf{k}}$ is not. We then use this exact solution to compute the rate of thermal photon emission in the QGP, probing whether non-hydrodynamic contributions to $f_{\mathbf{k}}$ leave an observable imprint.

\section{Kinetic theory setup}
\label{sec:kinetic-theory}
 
Bjorken flow~\cite{Bjorken:1982qr} assumes that the matter produced in a heavy-ion collision is boost-invariant along the beam axis, as well as homogeneous and azimuthally symmetric in the transverse plane. The features of such a flow profile are more conveniently captured adopting hyperbolic coordinates $(\tau, x, y, \eta_s)$, where the proper time, $\tau = \sqrt{t^2 - z^2}$, and the spacetime rapidity, $\eta_s = \frac{1}{2}\ln[(t+z)/(t-z)]$, replace the Cartesian $t$ and $z$ components. These assumptions imply that all fluid-dynamical quantities depend solely on $\tau$. Furthermore, the single-particle distribution function of a massless gas depends solely on $( \tau, k_0 (\tau), \cos\Theta(\tau) )$, with $\cos\Theta = k_\eta/(\tau k_0)$.

In the RTA, the Boltzmann equation takes the form
\begin{equation}
  \label{eq:boltzmann_rta}
  \partial_\tau f_{\mathbf{k}}
  = \frac{1}{\tau_R}
  \bigl( f_{0\mathbf{k}} - f_{\mathbf{k}} \bigr) \; ,
\end{equation}
where $\tau_R$ is the relaxation time, $f_{0\mathbf{k}} = \exp(\alpha - k_0/T)$ is the equilibrium distribution, $k_0$ is the particle's energy, $T$ is the temperature and $\alpha$ is the thermal potential. In this work, we assume the relaxation time is $\tau_R = 5\eta/(\varepsilon+P)$, with $\varepsilon$ being the energy density, $P$ the thermodynamic pressure and $\eta$ the shear viscosity. For the sake of simplicity, we assume a constant $\eta/s$, where $s$ is the entropy density.
 
The Boltzmann equation \eqref{eq:boltzmann_rta} can be solved exactly,
\begin{equation}
  \label{eq:fk_exact}
  f_{\mathbf{k}} (\tau) 
  = f_{\mathbf{k}} (\tau_0) 
  \mathcal{D} (\tau_0, \tau)  
  + \int_{\tau_0}^{\tau}
      \frac{d\tau'}{\tau_R(\tau')}
      \exp \left[ \alpha (\tau') - \frac{k_0 }{T(\tau')} \sqrt{ 1 + \cos^2 \Theta \left( \frac{\tau^2}{\tau'^2} - 1 \right)} \right]
      \mathcal{D} (\tau', \tau) \; ,
\end{equation}
where we defined $\mathcal{D} (\tau_1, \tau_2) = \exp\left[ -\int_{\tau_1}^{\tau_2} d \tau'/\tau_R(\tau') \right]$. This expression is exact within the RTA and avoids any expansion of $f_{\mathbf{k}}$ in its irreducible moments, thus circumventing the divergences associated with the moment expansion \cite{deBrito:2024qow}.
 
In the 14-moment approximation, on the other hand, $f_{\mathbf{k}}$ reads
\begin{equation}
  \label{eq:f-14mom}
  f_{\mathbf{k}}^{\text{14-moment}}
  = f_{0\mathbf{k}} \left[
    1 + \frac{k_0^2}{4T^2}\,P_2(\cos\Theta)\,\frac{\varrho_{2,1}}{\frac{3}{\pi^2} \, e^\alpha \, T^4} \right] \; ,
\end{equation}
where $P_n (x)$ denotes the Legendre polynomials.

\subsection{Determining thermodynamic quantities}
 
The solution~\eqref{eq:fk_exact} requires the knowledge of the dynamics of $T$ and $\alpha$. These are determined self-consistently from the irreducible moments of $f_{\mathbf{k}}$,
\begin{equation}
  \label{eq:def-moments}
  \varrho_{n,\ell}
  = \int \frac{d^3 \mathbf{k}}{(2 \pi)^3 k_0} \, k_0^n\, P_{2\ell}(\cos\Theta)\, f_{\mathbf{k}} \; ,
\end{equation}
which satisfy a hierarchy of coupled differential equations in Bjorken flow~\cite{Denicol:2021},
\begin{equation}
  \label{eq:hierarchy-eoms-bjorken}
  \frac{d \varrho_{n, \ell}}{d \tau} + \frac{1}{\tau_R} \left[ \varrho_{n, \ell} - e^\alpha \frac{(n+1)!}{2 \pi^2}T^{n+2} \delta_{\ell 0} \right] 
  + \mathcal{P}_{n, \ell} \frac{\varrho_{n, \ell - 1}}{\tau} 
  + \mathcal{Q}_{n, \ell} \frac{\varrho_{n, \ell}}{\tau} 
  + \mathcal{R}_{n, \ell} \frac{\varrho_{n, \ell + 1}}{\tau} = 0 \; ,
\end{equation}
where we have introduced the following coefficients,
\begin{subequations}  
  \begin{align}
    \mathcal{P}_{n, \ell} &= 2 \ell \frac{(n + 2\ell ) (2\ell - 1)}{(4\ell + 1) (4\ell -1 )} \; , \\
    \mathcal{Q}_{n, \ell} &=  \frac{2 \ell (2\ell + 1) + n (24\ell^2 + 12 \ell - 3)}{3 (4\ell - 1) (4\ell + 3)} + \frac{2}{3} \; , \\
    \mathcal{R}_{n, \ell} &= (n - 2 \ell -1) \frac{(2\ell + 1) (2\ell + 2)}{(4\ell + 1) (4\ell + 3)}. 
  \end{align}
\end{subequations}
We then solve Eqs.~\eqref{eq:hierarchy-eoms-bjorken} and fix the temperature and thermal potential by imposing Landau matching conditions, in which the out-of-equilibrium number and energy densities are fixed to their corresponding equilibrium counterparts, that is, $\varrho_{1,0} \equiv \frac{1}{\pi^2} \, e^\alpha \, T^3$ and $\varrho_{2,0} \equiv \frac{3}{\pi^2} \, e^\alpha \, T^4$.

\section{Hydrodynamization and thermalization}
\label{sec:hydro}

We compare two truncations of the hierarchy of moment equations \eqref{eq:hierarchy-eoms-bjorken}, in which all moments with $\ell > \ell_{\mathrm{max}}$ are set to zero: (i) the \textit{hydrodynamic solution}, in which we set $\ell_{\rm max} = 1$ and thus retain only the degrees of freedom that appear in the energy-momentum tensor, with $f_{\mathbf{k}}$ given by the 14-moment form~\eqref{eq:f-14mom}; and (ii) the \textit{exact solution}, obtained in the limit $\ell_{\rm max} \to \infty$ (in practice, we take $\ell_{\mathrm{max}} = 20$ since higher values do not appreciably change the results), with $T(\tau)$ and $\alpha(\tau)$ inserted into Eq.~\eqref{eq:fk_exact}, which is evaluated numerically without resorting to any moment reconstruction. In what follows, we shall always consider equilibrium initial conditions.

\subsection{Energy-momentum tensor}

In Bjorken flow, the energy-momentum tensor takes the form $T^{\mu\nu} = \mathrm{diag}(\varepsilon, P+\pi/2,\, P+\pi/2,\, P-\pi)$, where the shear-stress scalar $\pi = -\frac{2}{3}\varrho_{2,1}$ carries all the non-equilibrium information. 
\begin{figure}[ht]
  \centering
  \begin{subfigure}{0.32\textwidth}
    \includegraphics[width=\textwidth]{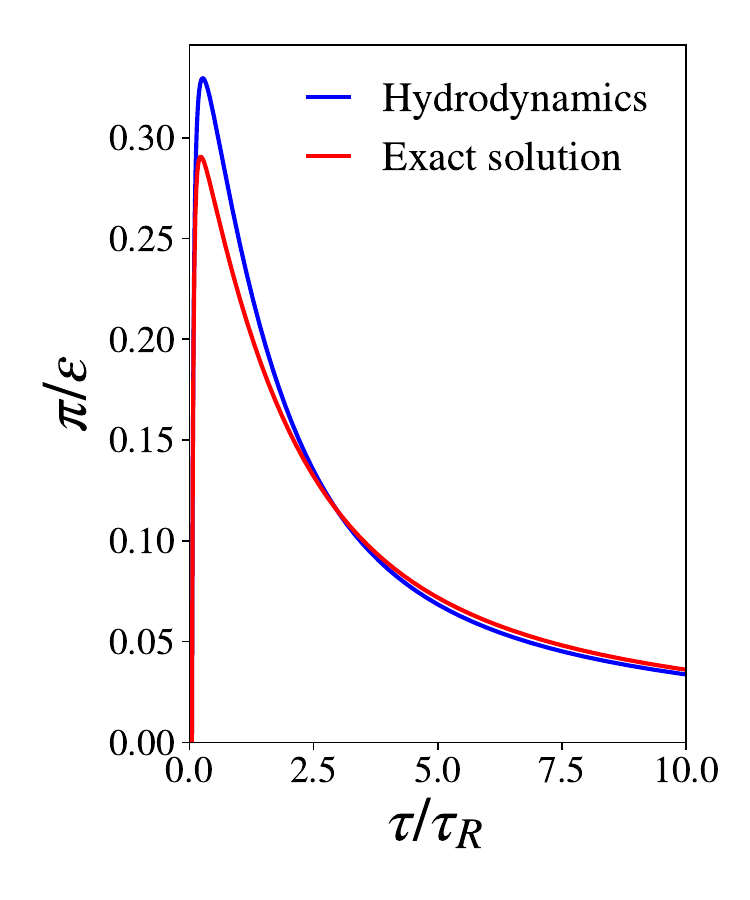}
  \end{subfigure}
  \begin{subfigure}{0.32\textwidth}
    \includegraphics[width=\textwidth]{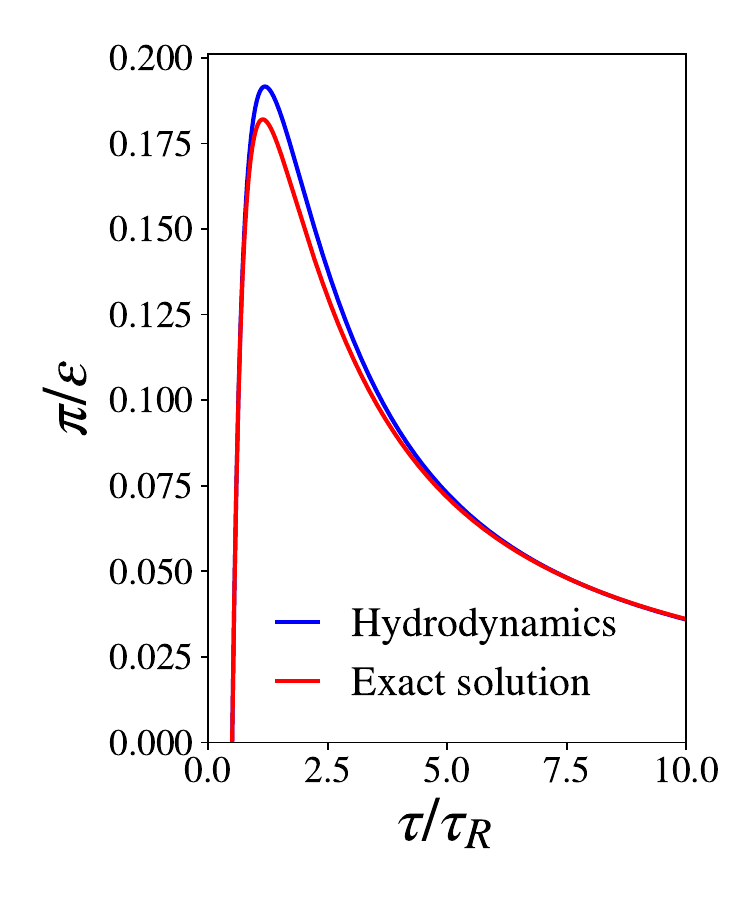}
  \end{subfigure}
  \begin{subfigure}{0.32\textwidth}
    \includegraphics[width=\textwidth]{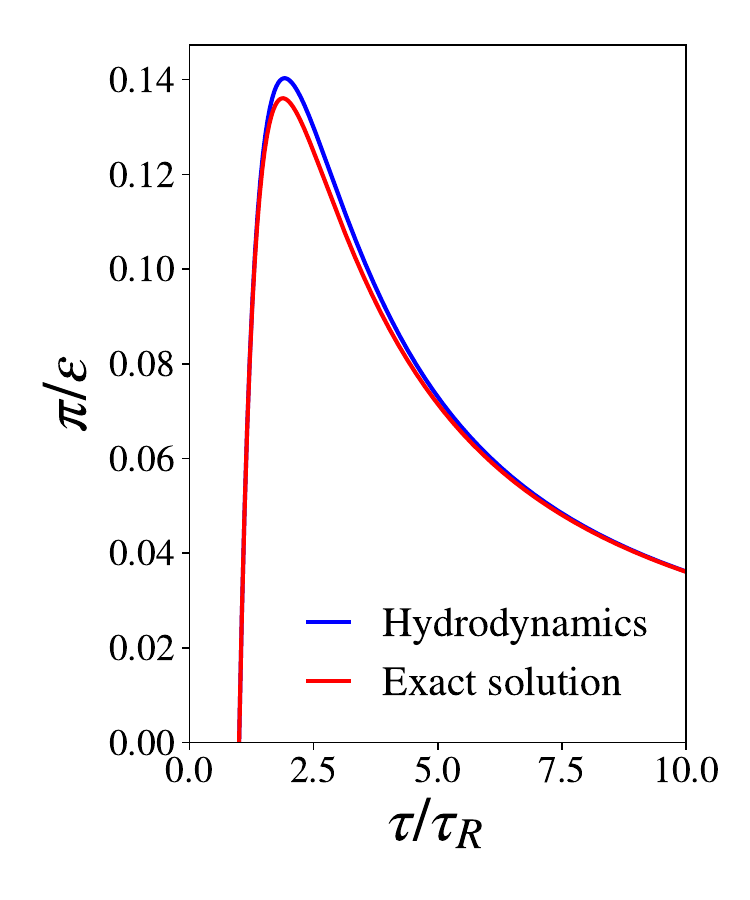}
  \end{subfigure}
  \caption{Shear-stress scalar rescaled by the energy density,
  $\pi/\varepsilon$, as a function of $\hat \tau$ considering $\hat\tau_0 = 0.05$ (left panel), $\hat\tau_0 = 0.5$ (central panel) and $\hat\tau_0 = 1.0$ (right panel). In all cases, we assume $\alpha(\hat \tau_0) = 0$ and $T (\hat \tau_0) = 200$ MeV.}
  \label{fig:Tmunu}
\end{figure}

Fig.~\ref{fig:Tmunu} shows the ratio $\pi/\varepsilon$ as a function of the \textit{rescaled} proper time $\hat\tau = \tau/\tau_R$ for three values of $\hat\tau_0$, namely $\hat\tau_0= 0.05$ (left panel), $\hat\tau_0 = 0.5$ (central panel) and $\hat\tau_0 = 1.0$ (right panel). We observe that the 14-moment (hydrodynamic) solution (blue curves) reproduces the exact solution (red curves) well, even when the system is considerably out of equilibrium. 

\subsection{Distribution function}

In the previous subsection, we have shown that a hydrodynamic truncation provides a good estimate for the energy-momentum tensor of a massless gas in Bjorken flow. The next step is to verify whether this further implies that such a system can be described by the 14-moment approximation on the microscopic level. 
\begin{figure}[ht]
  \centering
  \includegraphics[width=0.32\textwidth]{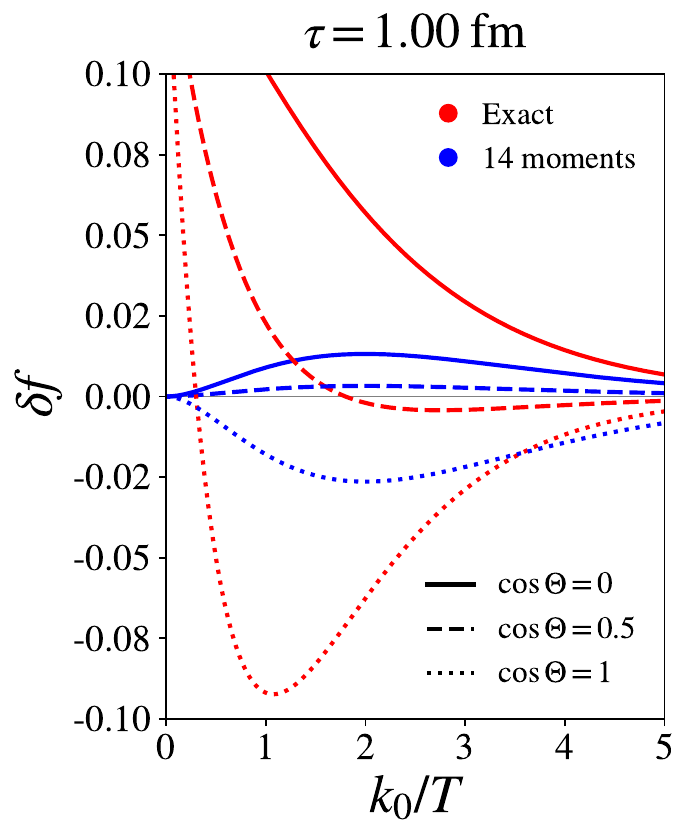}
  \includegraphics[width=0.32\textwidth]{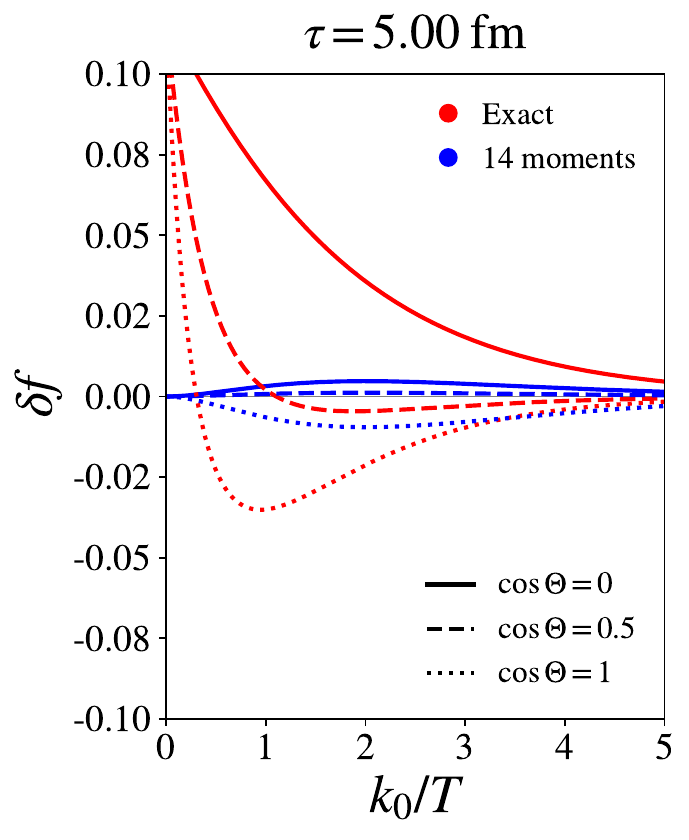}
  \includegraphics[width=0.32\textwidth]{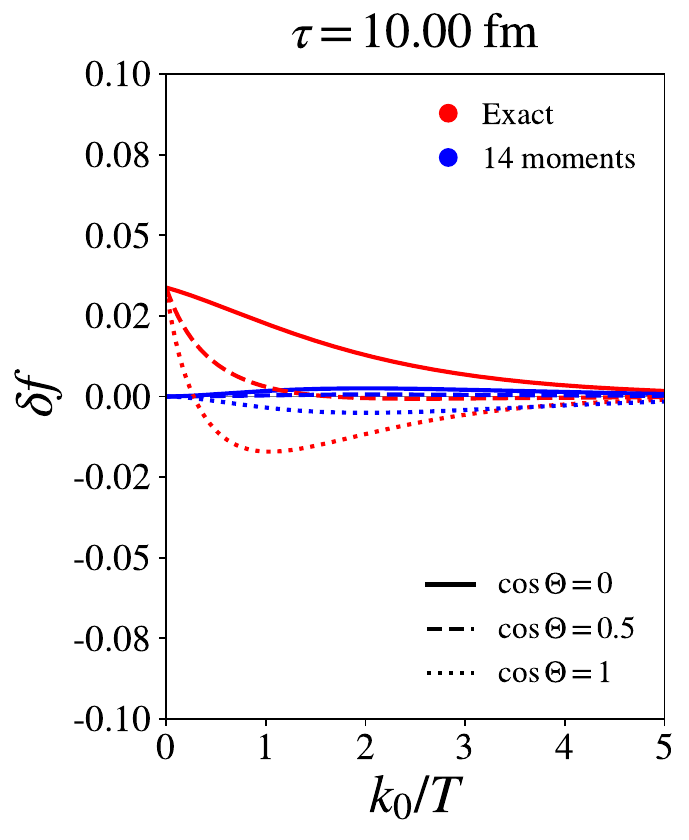}
  \caption{Exact (red curves) and hydrodynamic (blue curves) solutions of the non-equilibrium distribution function as a function of $k_0/T$ calculated at $\tau = 1.0$~fm (left panel), $\tau = 5.0$~fm (central panel) and $\tau = 10.0$~fm (right panel), for $\cos\Theta = 0$ (solid curves), $\cos\Theta = 0.5$ (dashed curves) and $\cos\Theta = 1$ (dotted curves). In all cases, we assume $\tau_0 = 0.5$ fm, $\eta/s = 0.2$, $\alpha(\tau_0) = 0$ and $T (\tau_0) = 200$ MeV.}
  \label{fig:delta-fk}
\end{figure}

In Fig.~\ref{fig:delta-fk}, we portray the non-equilibrium distribution function $\delta f_{\mathbf{k}}$ ($=f_{\mathbf{k}}-f_{0\mathbf{k}}$) as a function of $k_0/T$ for several values of $\cos\Theta$ computed at different times. We observe that the exact solution differs substantially from the 14-moment result, especially at early times. In particular, the fluid-dynamical truncation considerably underestimates the exact non-equilibrium corrections to $f_{\mathbf{k}}$, even when $T^{\mu\nu}$ is already well described by hydrodynamics.

\section{Thermal photon emission in the forward scattering approximation}
\label{sec:photon-emission-fsa}

Thermal photons in the QGP are predominantly produced through quark--anti-quark annihilation ($q\bar{q}\to\gamma g$) and Compton scattering ($qg\to\gamma q$,\,$\bar{q}g\to\gamma\bar{q}$)~\cite{Wong:1994}. In the forward scattering approximation (FSA), the photon is assumed to be emitted in the same direction as the incoming quark or anti-quark. In this case, the emission rate is given by \cite{Dusling:2010}
\begin{equation}  
  \label{eq:FSA}
  k_0\frac{dN}{d \mathbf{k}}
  = \frac{5}{9} \, \frac{\alpha_e \, \alpha_s}{2\pi^2} \,
  f^{(q)}_{\mathbf{k}} \, T^2 \, \ln \left( \frac{3.7388 \, k_0}{g_s^2 \, T} \right) \; ,
\end{equation}
where $N$ is the number of photons per unit four-volume, $f^{(q)}_{\mathbf{k}}$ is the quark distribution function, $\alpha_e$ is the electromagnetic fine-structure constant and $\alpha_s = g_s^2/(4\pi)$, with $g_s$ being the strong coupling. We approximate $f^{(q)}_{\mathbf{k}}$ by the distribution obtained from the kinetic theory calculation detailed in the previous section. We remark that the standard approach in the literature assumes the 14-moment approximation for $f^{(q)}_{\mathbf{k}}$, see e.g., Ref.~\cite{Paquet:2016}. In this work, we quantify how non-equilibrium corrections beyond the hydrodynamic limit affect the thermal photon emission rate.

\begin{figure}[ht]
  \centering
  \includegraphics[width=0.32\textwidth]{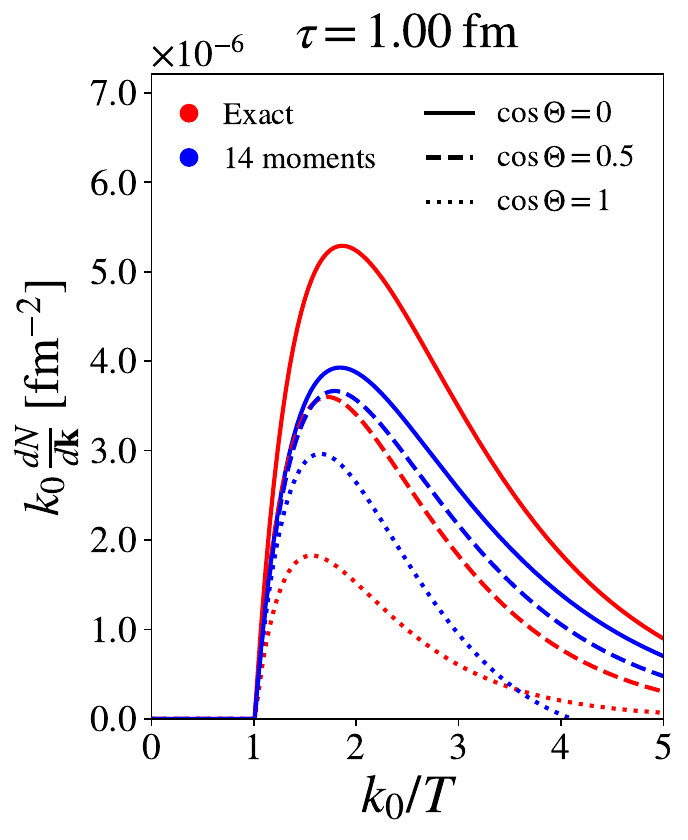}
  \includegraphics[width=0.32\textwidth]{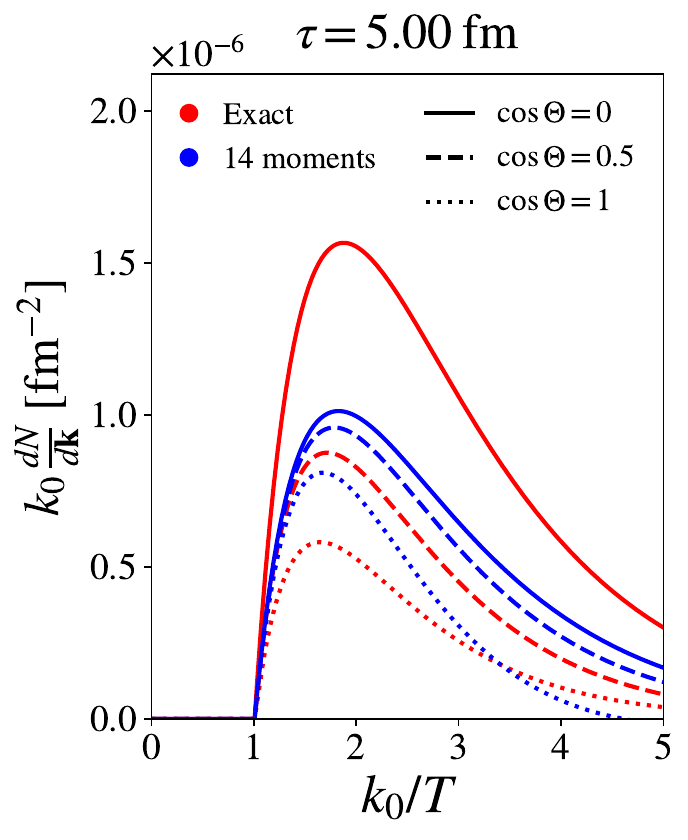}
  \includegraphics[width=0.32\textwidth]{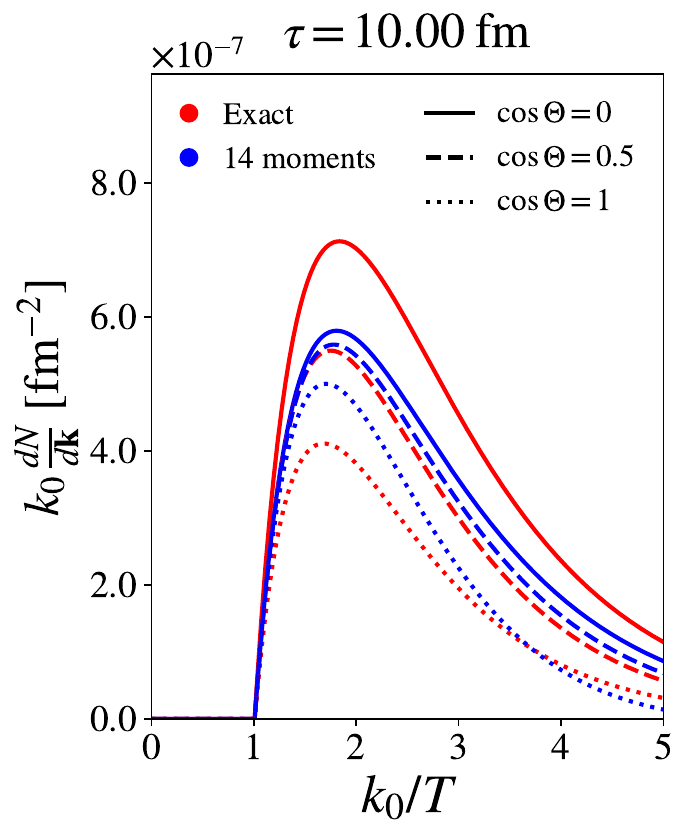}
  \caption{Exact (red curves) and hydrodynamic (blue curves) solutions of the thermal photon emission rate as a function of $k_0/T$ calculated at $\tau = 1.0$~fm (left panel), $\tau = 5.0$~fm (central panel) and $\tau = 10.0$~fm (right panel), for $\cos\Theta = 0$ (solid curves), $\cos\Theta = 0.5$ (dashed curves) and $\cos\Theta = 1$ (dotted curves). In all cases, we assume $\tau_0 = 0.5$ fm, $\eta/s = 0.2$, $\alpha(\tau_0) = 0$, $T (\tau_0) = 200$ MeV and $\alpha_s = 0.3$.}
  \label{fig:photon-FSA}
\end{figure}
The thermal photon emission rate in the forward scattering approximation is displayed in Fig.~\ref{fig:photon-FSA} as a function of $k_0/T$ at different times and considering different values of $\cos\Theta$. We compare the exact and 14-moment solutions for $f^{(q)}_{\mathbf{k}}$, given by Eqs.~\eqref{eq:fk_exact} and \eqref{eq:f-14mom}, respectively. We observe that non-equilibrium effects beyond the hydrodynamic limit considerably affect the magnitude and angular dependence of the thermal photon production. In particular, the exact distribution makes the local photon emission more anisotropic than the 14-moment approximation predicts: the exact rate is enhanced at $\cos\Theta=0$ and suppressed as $\cos\Theta$ is increased. This result suggests that non-equilibrium effects beyond the hydrodynamic limit are physically relevant for photon production.

\section{Summary and conclusions}
\label{sec:summary}

We have studied the dynamics of a classical gas of massless particles undergoing Bjorken expansion in the context of kinetic theory. We showed that the energy-momentum tensor of such a system is well described by hydrodynamics, even though the 14-moment approximation fails to capture the full non-equilibrium structure of the single-particle distribution function. This effect was then quantified by computing the photon emission rate in the forward scattering approximation, in which case we observed that the inclusion of non-hydrodynamic contributions might be essential in order to obtain a quantitatively accurate description. Non-equilibrium corrections beyond the hydrodynamic limit make the local emission rate considerably more anisotropic, enhancing it at $\cos \Theta=0$ and suppressing it at larger $\cos \Theta$, which hints at a potential observational handle on the thermalization state of the QGP.

\acknowledgments

C.V.P.B.\ is supported by the Research Council of Finland Center of Excellence in Quark Matter, Project No.\ 364192, and the European Research Council, Project No.\ ERC-2018-ADG-835105 YoctoLHC. G.S.D.\ is supported by Conselho Nacional de Desenvolvimento Científico e Tecnológico through Grant No.~307761/2022-3 and acknowledges the support of the INCT-FNA, Grant No.~408419/2024-5.



\begin{thebibliography}{99}

\bibitem{Heinz:2013th}
  U.~Heinz and R.~Snellings,
  \href{https://doi.org/10.1146/annurev-nucl-102212-170540}{Ann.\ Rev.\ Nucl.\ Part.\ Sci.\ \textbf{63}, 123 (2013)}.

\bibitem{Bjorken:1982qr}
  J.~D.~Bjorken,
  \href{https://doi.org/10.1103/PhysRevD.27.140}{Phys. Rev. D \textbf{27}, 140--151 (1983)}.
  
\bibitem{Anderson:1974nyl}
  J.~L.~Anderson and H.~R.~Witting,
  \href{https://doi.org/10.1016/0031-8914(74)90355-3}{Physica \textbf{74}, no.\ 3, 466--488 (1974)}.

\bibitem{deBrito:2024qow}
  C.~V.~P.~de Brito, D.~Wagner, G.~S.~Denicol and D.~H.~Rischke,
  \href{https://doi.org/10.1103/29mk-wwkg}{Phys. Rev. D \textbf{113}, 096007 (2026)}.
  
\bibitem{Denicol:2021}
  G.~S.~Denicol and D.~H.~Rischke,
  \emph{Microscopic Foundations of Relativistic Fluid Dynamics},
  Springer (2021).

\bibitem{Wong:1994}
  C.-Y.~Wong,
  \emph{Introduction to High-Energy Heavy-Ion Collisions},
  World Scientific (1994).

\bibitem{Dusling:2010}
  K.~Dusling,
  \href{https://doi.org/10.1016/j.nuclphysa.2010.04.001}{Nucl.\ Phys.\ A \textbf{839}, 70 (2010)}.

\bibitem{Paquet:2016}
  J.-F.~Paquet, C.~Shen, G.~S.~Denicol, M.~Luzum,
  B.~Schenke, S.~Jeon and C.~Gale,
  \href{https://doi.org/10.1103/PhysRevC.93.044906}{Phys.\ Rev.\ C \textbf{93}, 044906 (2016)}.




\end{thebibliography}
\end{document}